\documentclass{sigchi}

\toappear{\scriptsize Permission to make digital or hard copies of all or part of this work for personal or classroom use is granted without fee provided that copies are not made or distributed for profit or commercial advantage and that copies bear this notice and the full citation on the first page. Copyrights for components of this work owned by others than ACM must be honored. Abstracting with credit is permitted. To copy otherwise, or republish, to post on servers or to redistribute to lists, requires prior specific permission and/or a fee. Request permissions from permissions@acm.org. \\
{\emph{ASSETS '19, October 28--30, 2019, Pittsburgh, PA, USA.} } \\
Copyright \copyright~2019 Association of Computing Machinery. \\
ACM ISBN 978-1-4503-6676-2/19/10\ ...\$15.00. \\
http://dx.doi.org/10.1145/3308561.3353790}

\usepackage{changepage}

\usepackage{balance}       
\usepackage{graphicx}      
\usepackage[T1]{fontenc}   
\usepackage{txfonts}
\usepackage{mathptmx}
\usepackage[pdflang={en-US},pdftex]{hyperref}
\usepackage{color}
\usepackage{booktabs}
\usepackage{textcomp}

\usepackage{microtype}        
\usepackage{ccicons}          

\def\plaintitle{3D Printed Maps and Icons for Inclusion: Testing in the Wild by People who are Blind or have Low Vision}
\def\plainauthor{Leona Holloway, Kim Marriott,  Matthew Butler, Samuel Reinders}
\def\plainkeywords{maps; blind; low vision; vision impairment; 3D printing; orientation \& mobility}

\makeatletter
\def\url@leostyle{%
  \@ifundefined{selectfont}{
    \def\UrlFont{\sf}
  }{
    \def\UrlFont{\small\bf\ttfamily}
  }}
\makeatother
\def\pprw{8.5in}
\def\pprh{11in}

\definecolor{linkColor}{RGB}{6,125,233}
\hypersetup{%
  pdftitle={\plaintitle},
  pdfauthor={\plainauthor},
  pdfkeywords={\plainkeywords},
  pdfdisplaydoctitle=true, 
  bookmarksnumbered,
  pdfstartview={FitH},
  colorlinks,
  citecolor=black,
  filecolor=black,
  linkcolor=black,
  urlcolor=linkColor,
  breaklinks=true,
  hypertexnames=false
}

\begin{document}

\title{\plaintitle}

\numberofauthors{4}
\author{%
  \alignauthor{Leona Holloway\\
    \affaddr{Monash University}\\
    \affaddr{Melbourne, Australia}\\
    \email{Leona.Holloway@monash.edu}}\\
  \alignauthor{Kim Marriott\\
    \affaddr{Monash University}\\
    \affaddr{Melbourne, Australia}\\
    \email{Kim.Marriott@monash.edu}}\\
  \alignauthor{Matthew Butler\\
    \affaddr{Monash University}\\
    \affaddr{Melbourne, Australia}\\
    \email{Matthew.Butler@monash.edu}}\\
  \alignauthor{Samuel Reinders\\
    \affaddr{Monash University}\\
    \affaddr{Melbourne, Australia}\\
    \email{Samuel.Reinders@monash.edu}}\\
}

\maketitle
\begin{abstract}
The difficulty and consequent fear of travel is one of the most disabling consequences of blindness and severe vision impairment, affecting confidence and quality of life. Traditional tactile graphics are vital in the Orientation and Mobility training process, however 3D printing may have the capacity to enable production of more meaningful and inclusive maps. This study explored the use of 3D printed maps on site at a public event to examine their suitability and to identify guidelines for the design of future 3D maps. An iterative design process was used in the production of the 3D maps, with feedback from visitors who are blind or have low vision informing the recommendations for their design and use. For example, it was found that many representational 3D icons could be recognised by touch without the need for a key and that such a map helped form mental models of the event space. Complex maps, however, require time to explore and should be made available before an event or at the entrance in a comfortable position. The maps were found to support the orientation and mobility process, and importantly to also promote a positive message about inclusion and accessibility.
\end{abstract}




\def\plainkeywords{Maps; Blind; Low Vision; Vision Impairment; 3D Printing; Orientation \& Mobility}


\section{Introduction}
People who are blind or have low vision (BLV) consistently identify independent travel as amongst their greatest challenges~\cite{Engelstad2019, Sheffield2016}, affecting confidence, independence and quality of life~\cite{Keeffe2005}. 
Orientation and Mobility (O\&M) training is widely used to address this barrier to independent living by equipping BLV people with the  skills for safely moving around the environment and by familiarising  them  with their local environment and new locations and routes before travel. Raised line drawings, called \emph{tactile graphics},  are commonly used as part of this process, with tactile maps helping the viewer to build up a mental model  of the geography prior to venturing forth~\cite{rowell2005,ungar1993}.  However, the provision of tactile maps is limited because they are a specialised format used by a small number of people and because they require considerable experience to use and understand~\cite{Gardner1996, Holloway2016}.

As an alternative to tactile maps, handmade  bespoke 3D models are occasionally provided for the use of visitors, including BLV people, at significant cultural sites (Figure~\ref{fig:figure2}).  At present, their provision is limited by the difficulty and cost of production. Commodity 3D printers are set to change this, enabling the production of 3D models at comparable effort and price to tactile graphics. A number of  studies have suggested the use of 3D printed models in O\&M training~\cite{celani2007,gual2011,gual2012analysis,Holloway2018printed, taylor2015tactilemaps, voigt2006development}. In our previous work, we found that 3D printed models were strongly preferred over tactile graphics as they were better at showing relative height and more memorable~\cite{Holloway2018printed}. It was also found that understanding was enhanced by the use of 3D printed icons~\cite{Holloway2018printed}, i.e. representational 3D symbols with a physical resemblance to the object or concept they are representing~\cite{Robinson1984}.  
 However, to the best of our knowledge there has been no evaluation of the use of 3D models for O\&M training ``in the wild.'' Testing in natural settings has gained popularity in the HCI field in recognition of the importance of external factors that influence the way technology is used~\cite{Rogers2017}. ``In the wild'' testing is the key novelty of the current paper.
 
\begin{figure*}
\centering
  \includegraphics[width=\textwidth]{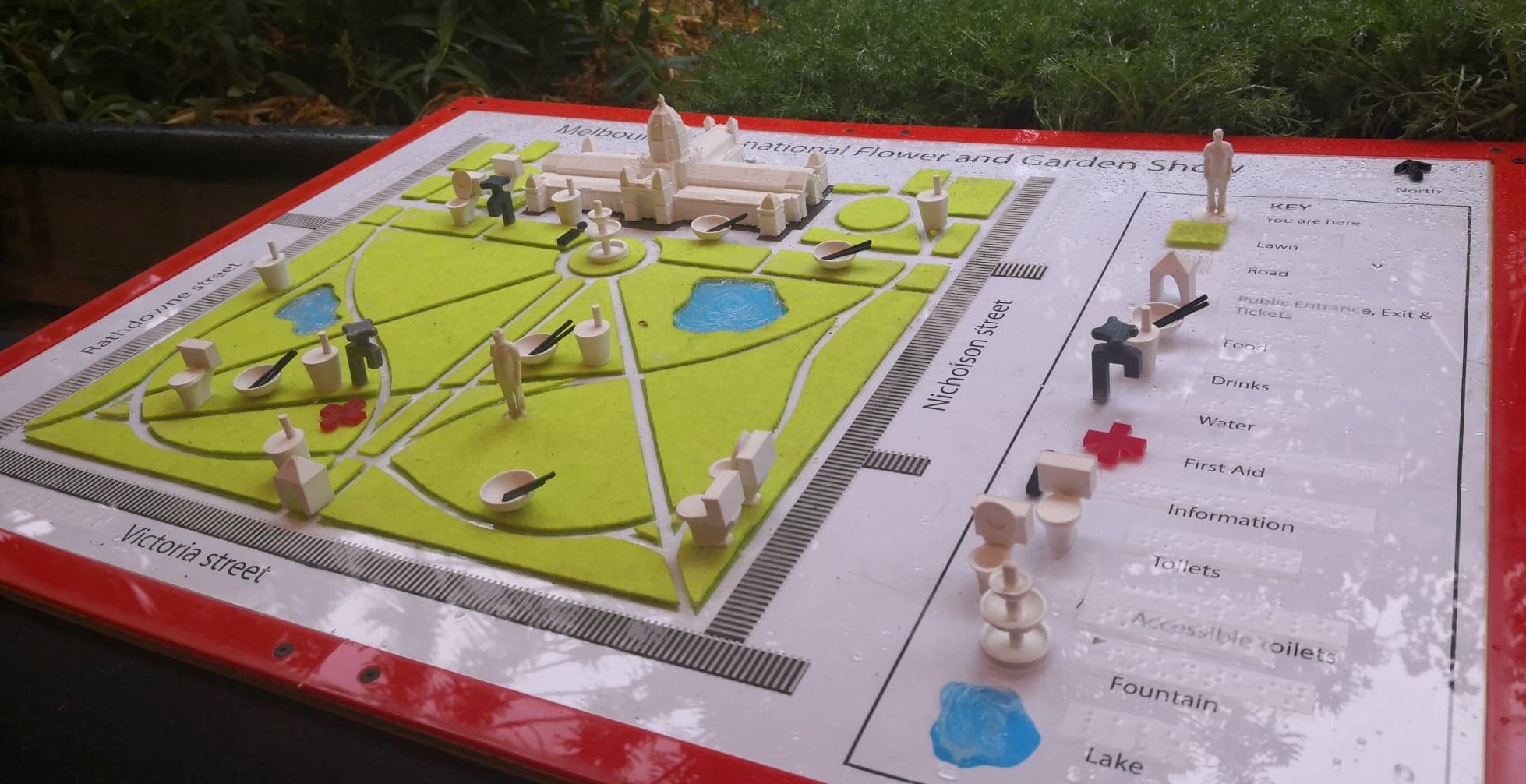}
  \caption{Large map of the Melbourne International Flower and Garden Show on site.}~\label{fig:figure1}
\end{figure*}

Our study is the result of a collaboration with a not-for-profit O\&M training organisation, Guide Dogs Victoria (GDV), to provide 3D printed models for BLV visitors to the 2019 Melbourne International Flower and Garden Show. The Garden Show takes place annually over five days with hundreds of stands and exhibits in the South Carlton Gardens, a cultural landmark covering 31 acres.  An estimated 103,000 people attended the Garden Show in 2019. GDV hosted a Sensory Garden and sponsored a competition Show Garden. As part of an effort to create an inclusive experience for BLV visitors, GDV requested accessible 3D maps of the event, which formed the basis of this study.

The study aimed to investigate two fundamental research questions:

\begin{itemize}
  \setlength\itemsep{0em}
\item [RQ1.] \emph{What guidelines should be used for the creation of 3D maps?} One particular focus was the design of 3D printed map icons for immediate recognition by touch. We also examined issues of scale, texture, realism and complexity. 

\item [RQ2.]	\emph{What role can 3D maps play in O\&M for events such as the Garden Show?} We particularly wished to find out whether the provision of 3D maps at the event itself was useful, whether  a smaller portable 3D printed map might be helpful for wayfinding while navigating the event, and whether the maps could be used without prior training and experience. 

\end{itemize}

In preparation for the study we collaborated with GDV in the design and presentation of four maps for BLV visitors to the Garden Show. These maps were made available by GDV at the Garden Show and feedback was gathered from ten BLV visitors and four sighted GDV representatives. 

The key findings include:  
\begin{itemize}
  \setlength\itemsep{-0em}
    \item Representational 3D icons can be readily recognised by touch, even without experience or confidence using tactile maps. However, labelling is still required to support understanding. (RQ1)
    \item When designing 3D icons it is important to leave space  between adjacent icons for touch reading and to place  the most recognisable features at the top, where they can most easily be accessed by touch. (RQ1)
  \item 3D  maps  support orientation and mobility through identification of landmarks, route planning and creation of a mental map in real life situations. (RQ2)
    \item The environment in which a map is provided is very important. Complex 3D maps require time to explore by touch and should be made available in a relaxed environment and appropriate location, such as a site entrance or even at home prior to visiting the location. (RQ2)
    \item 3D maps provide a positive message about inclusion and accessibility and are of interest and use to the sighted community as well as the BLV community. (RQ2)
\end{itemize}

\subsection{Contributions}
This study makes a number of contributions to the area of accessible materials production. It is the first ``in the wild'' study of 3D maps for BLV people. It is also the first study of representational 3D icons for touch reading. As such, the study has significant implications for O\&M training, informing the creation of professional guidelines for the use and design of accessible 3D prints, as well as clarifying the potential role of 3D maps for accessibility and inclusion.

\section{Related work}

\subsection{Tactile Maps}
Accessibility guidelines recommend the use of  tactile graphics to convey graphical content such as maps and plans to BLV people~\cite{edman1992tactile, BANA2010}. Tactile graphics and maps can be produced using a braille embosser, printing onto microcapsule swell paper or thermoforming~\cite{rowell2003world}. However, embossing has low fidelity, microcapsule paper is unable to produce elevations higher than 0.5mm, and thermoforming is unable to create sharp vertical features or overhangs.

\begin{figure}
\centering
  \includegraphics[width=\columnwidth]{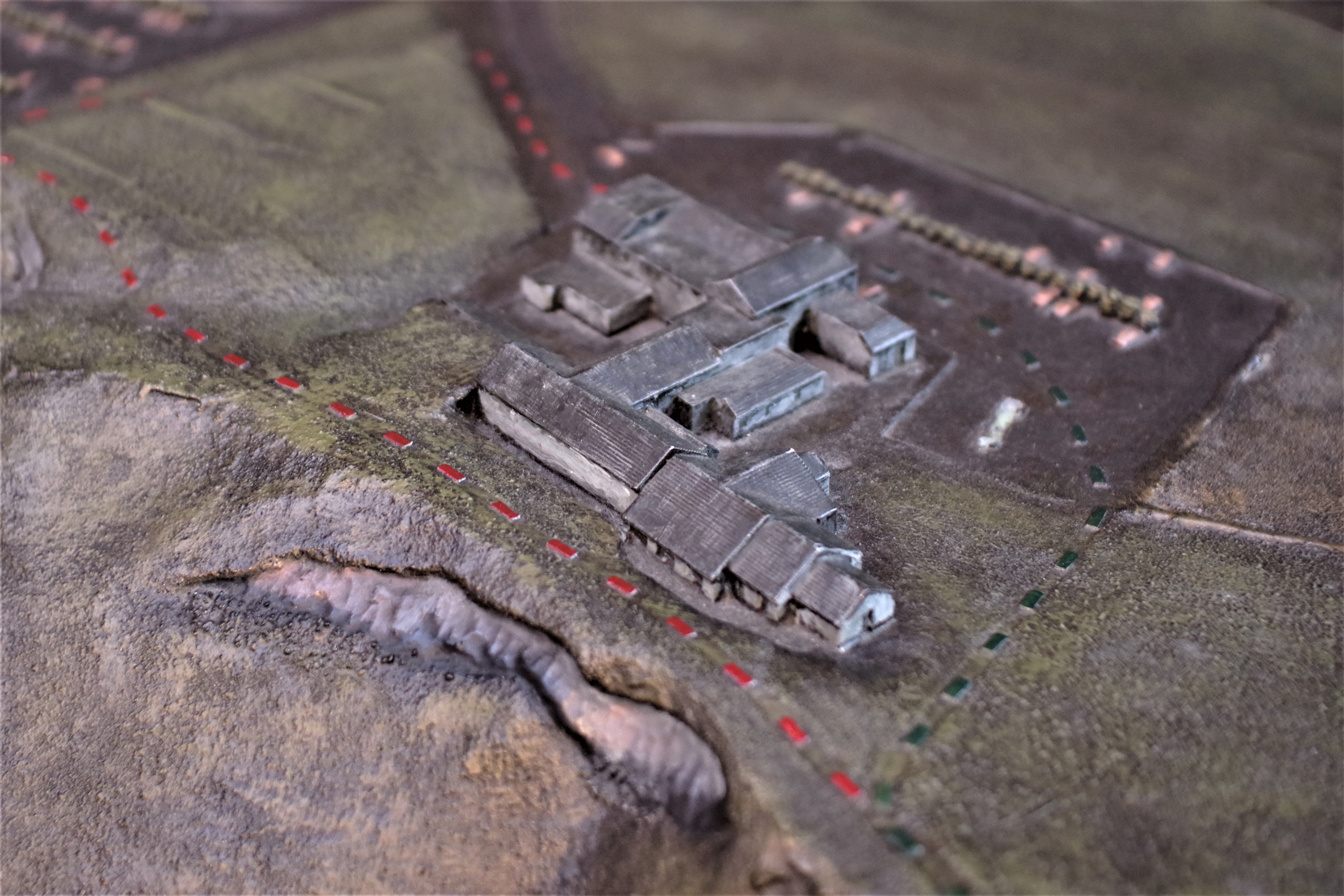}
  \caption{Bespoke 3D model of the visitor center and walking trails at the Giant's Causeway, Ireland. }~\label{fig:figure2}
\end{figure}

3D models logically seem better suited to showing information that is three-dimensional in nature. Bespoke 3D models typically cast in metal can be found at many European cultural landmarks (see Figure~\ref{fig:figure2}). Designed to provide BLV people with a greater sense of their surrounding environment~\cite{reliefstrasbourg}, these 3D maps have also been embraced by the general public~\cite{reliefgermany}. In practice, such 3D models are relatively rare due to difficulties and cost of production and distribution. 

The advent of low cost digital  manufacturing technologies, such as 3D printing or laser cutting, has spawned interest in how these technologies can be used to create accessible maps and other kinds of graphics~\cite{siu20143D}. G\"{o}tzelmann~\cite{Gotzelmann2016} investigated the combination of 3D printed tactile overlays  with multi-touch, Taylor et al.~\cite{taylor2015tactilemaps} developed a web-based tool capable of generating 3D printable maps, and Giraud et al.~\cite{Giraud2017} embedded audio descriptions into 3D printed maps and found that this enhancement increased spatial memorisation by BLV people.

\subsection{Use in O\&M}
A primary use of tactile maps is in O\&M training. They allow BLV people to develop mental models of new spaces and environments prior to visiting~\cite{rowell2005,ungar1993}. Studies show that BLV children and adults can benefit from tactile maps, allowing them to develop spatial knowledge relating to the layout of environments~\cite{blades1999map,Ungar1994} and route wayfinding~\cite{Espinosa1998,Gardiner2003}. 
Additionally, when tactile maps are used in combination with direct experience, they can facilitate the development of richer environmental and route knowledge, outperforming that gained solely through direct experience~\cite{Espinosa1998,ungar2000,ungar1993}. 

Digitally produced 3D models may have additional advantages when used in O\&M training. Voigt and Martins~\cite{voigt2006development} suggested that 3D printed models of buildings should be used for O\&M training. Celani and Milan~\cite{celani2007} found that BLV people were positive towards 3D printed floor plans. 3D printed maps have also been used to build knowledge of cultural sites~\cite{Rossetti2018}. Gual et al.~\cite{gual2011} found that 3D printed maps were a useful tool in helping BLV people learn environments and helped in the identification and memorisation of routes, with additional work confirming these findings~\cite{gual2012analysis,Gual2015}. Our previous work comparing 3D printed maps with tactile equivalents found that 3D printed maps were highly preferred. The comparison highlighted that the 3D printed maps allowed more easily understandable icons to be used, for the height of elements to be more accurately conveyed, and that their use facilitated better short-term recall~\cite{Holloway2018printed}. 

To date, the majority of research in this area has used simulated or abstracted environments rather than real life situations~\cite{Perkins2002}. To the best of our knowledge, our study is first to  investigate the use of digitally produced 3D maps for O\&M in a real-world situation.

\subsection{Map \& Icon Design}
Tactile graphics production and design has a long history~\cite{Eriksson1999}. Tactile maps generally consist of a series of raised features composed of flat point, area or line symbols~\cite{Giraud2017,Lobben2012}, together with braille labels and descriptions. The content included in tactile maps is usually simplified and the features are selected for easy discernment and memorability. This is because haptic perception has a much lower resolution compared to visual perception~\cite{Jacobson1998,rowell2003world}.

User studies have been conducted to elicit information requirements for tactile maps, outlining the major features that BLV people want tactile maps to include~\cite{Papdopoulos2016,rowell2005}. Banovic et al.~\cite{Banovic2013} focused on investigating the different types of spatial information that BLV people need when learning and exploring environments: Generalised descriptions of the area, safety and navigation information, and places of interest. These broad categorisations closely align with the work undertaken by Papadopoulos et al.~\cite{Papdopoulos2016} suggesting that the most significant features needed in tactile maps include entrances/exits, toilets and indications of hazardous areas. Rowell and Ungar surveyed both BLV people~\cite{rowell2005} and tactile  graphic producers~\cite{rowell2003world} to investigate tactile map design and requirements, identifying the most important map features as roads and pathways, transport hubs, entrances, braille labels with a key, and orientation cues. 

Braille labels have traditionally been an essential component of tactile graphics and maps and are the only means of providing labels for people who are deafblind. However, with the majority of blindness caused by age-relate disease, as few as 10\% of blind people may be braille literate~\cite{NFIB2009}, and due to the space that braille labelling occupies, braille labelling typically utilises complex legends, increasing cognitive load. For this reason,
the use of audio labels  and multi-modal maps is an active field of research~\cite{baker2014tactile,Giraud2017,shi2016tickers}. However, this is still an area of much experimentation and is unlikely to be adopted by accessible format providers in the near term. Our study instead focuses on addressing some of the core design issues of 3D printed maps and in particular to what extent 3D printing allows the creation of more representative icons that can be identified without labels. 

The number of symbols and icons that can be reasonably included in a tactile map is limited, with studies recommending that no more than 10-15 distinct symbols be used in order to ensure they can be discerned from one another~\cite{Lobben2012,rowell2003world}. Unlike in the cartographic and geospatial design space, where universal standardised sets of symbols have been well defined, there exists no defined standard for tactile map iconography. This may be in part because the capabilities and needs of BLV people vary so widely~\cite{rowell2005}. Efforts into standardised design have been made~\cite{goodrick1985national,Lobben2012,rowell2003taxonomy}, but larger attempts to define a lexicon of standardised tactile symbols and textures for tactile maps have repeatedly failed due to the small number that can be distinguished from one another on a single map~\cite{Eriksson1999}. 

A further limitation of the tactile symbols used on tactile maps is that they are typically abstract. This contrasts with the symbols used in visual maps which are often iconic, visually resembling the object they represent, (e.g. a tree) some associated object (e.g. a pick axe for a mine) or an emblem associated with the object (e.g. red cross for hospital)~\cite{maceachren2004maps}. Such  representational symbols can often be understood without recourse to a legend and are more memorable than abstract symbols~\cite{Forrest1985,Leung2002}. 

3D models allow the use of volumetric icons and  may assist in the development of a wider lexicon of symbols. Research into the use of volumetric icons suggests that they should be used alongside more traditional flat symbols. One such comparison study found that volumetric symbols were preferred for representing stairs but flat symbols were favoured for directional arrows~\cite{McCallum2006}, while we found that volumetric symbols were useful when representing three-dimensional landmarks and features~\cite{Holloway2018printed}. Gual et al.~\cite{gual2012analysis} explored the use of abstract volumetric symbols, produced using 3D printing, and found that if symbol design was simple enough, designs were able to be recognised and discerned. Subsequent research involving BLV people found an improvement in discrimination of symbols in a legend if volumetric symbols were mixed with 2D symbols~\cite{Gual2014}, and also noted improvements in speed and accuracy when finding symbols on a tactile floor plan~\cite{Gual2015}. 

A major contribution of our study is to investigate the design and understandability of more representational 3D map icons ``in the wild."

\section{Design of maps}
The Garden Show offered an ideal opportunity to test the use of 3D maps in a realistic event setting with a reasonable number of expected BLV visitors. GDV made every effort to create an inclusive experience, installing beacons throughout the Garden Show for navigation using the BlindSquare Events app. Their Sensory Garden included scented plants, plants with interesting textures, auditory cues, braille labels, and varied ground textures. As an additional means of inclusion and access, GDV requested three accessible 3D maps: A large map of the Carlton Gardens with key features marked; a model of the GDV Sensory Garden; and a model of the Rob Waddell Show Garden. A fourth map was created for research purposes, this being a small hand-held map of the Carlton Gardens.

The maps of the grounds were intended to provide an accessible version of the print map that was available on the Garden Show website and being handed to all visitors at the gate. The accessible version could give an overview of the layout of the grounds, indicate what is available at the event, and assist with route planning. 
The models of the two gardens were intended to provide an overview of the spaces, and to provide information about those areas of the gardens where entry was not permitted: A corner of the Sensory Garden with a compost sculpture and all of the Show Garden. 
The mini map was designed to serve as a tactile reminder of the layout of the Gardens after visitors had studied the larger map. They would be able to mark points of interest with tactile stickers, take the mini map with them and refer to it when required. 

The maps were designed and constructed by a researcher with 20 years of experience in tactile graphics production. Throughout the design phase, GDV staff were afforded the opportunity to provide feedback.

\subsection{Large map of the Carlton Gardens}
The original print map was extremely complex, with 13 different icons used 31 times and over 100 site numbers. However, tactile maps need to be simple to allow ease of use~\cite{Gardiner2002,Gardner1996}. In consultation with representatives from GDV, it was decided that the most important map features were: Entry/exit points, the Royal Exhibition Building, roads, paths, grass, lakes, toilets, information booth, and a north indicator. Other features that should be included if possible were: Toilet accessibility, tram stops, first aid, food/beverages, coffee, water, and ``you are here''. The remaining features could be omitted: Trees, no entry, site numbers, media tent, event management, parents' room, and text labels. These choices align with research findings that the features BLV people most value on accessible maps are general information about the area, safety and navigation information, and points of interest~\cite{Banovic2013,Papdopoulos2016}. 

Icons were ideally to be printed at a maximum width of 1.5cm. Ideas for meaningful icons were sought through the Australia \& New Zealand Accessible Graphics Group, with suggestions provided by five touch readers and two experienced accessible formats producers. Some visual conventions were thought to be well understood by BLV people (e.g. arrows and a red cross for first aid) but others were not (e.g. water droplet). Potential icons were then pre-tested by sighted participants to determine which could be identified most easily by non-expert touch. Icons were rejected if they could not be readily identified or if their meaning was unclear. For example, the knife and fork on a dinner plate could not be distinguished tactually and the hamburger could easily be misinterpreted as a cushion or bush (Figure 3). 

\begin{figure*}
\centering
  \includegraphics[width=0.75\textwidth]{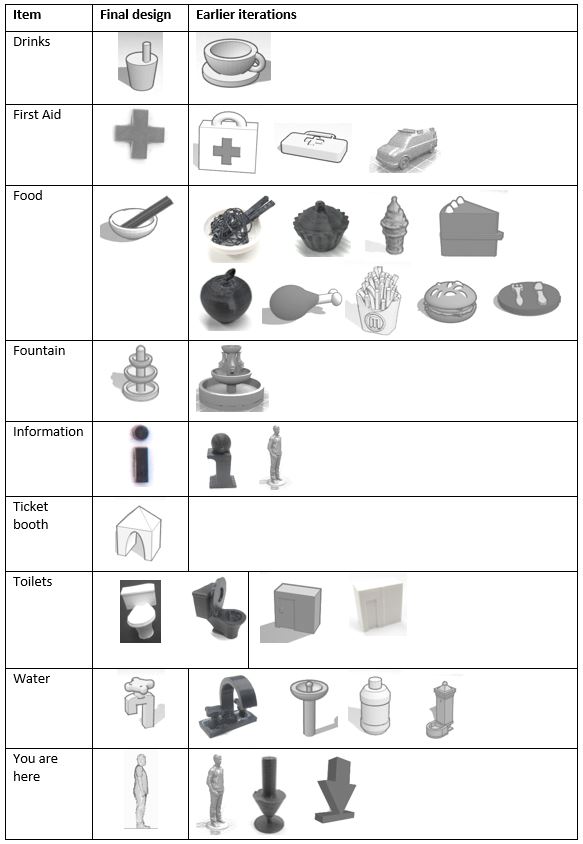}
  \caption{3D printed or laser cut icons used on the large map of the Carlton Gardens, along with earlier iterations.}~\label{fig:figure3}
\end{figure*}

Most of the selected icons were not direct translations of the original visual icons, but had instead been re-designed. In general, it was found that icons could most easily be recognised tactually if they were very simple, related to something that is usually touched, of a distinctive shape, and had the important features on the top of the model. Most icons were printed at 1.5--2.5cm in width and/or height. The "you are here" indicator was deliberately made the tallest icon on the map at 4cm high so that it would be the easiest to find.

The Carlton Gardens include the historic landmark Royal Exhibition Building. A pre-existing 3D printable model of the Building was provided by Heritage Victoria. This model was printed to scale with the Gardens and included a lot of detail such as towers and doorways.

The overall map was originally produced at A3 size. While all of the icons fit onto the map, there was very little space between them, making it more difficult to feel the sides and therefore recognise the shapes. The final model (Figure 1) was produced at A2 size to allow more space for exploration with the fingers. A minimum of 1cm gap was allowed between all adjacent icons. 

The base was produced from laser-cut clear acrylic, with laser etching to indicate roads. Green acrylic felt of 1mm thickness was laser cut and adhered to indicate the lawn areas. Paths were cut out of the felt to best represent the kerb and step down from the lawn to the path, with a minimum path width of 7mm. A clear print map was placed under the acrylic. In keeping with clear print guidelines~\cite{UKAAF2012, RT2011}, a sans serif font (Myriad Pro) was used at 55 point for the heading, 36 point for the road names and 21 point for the key. Clear braille labels were affixed alongside the print. A black base area was printed under the Exhibition Building to provide visual contrast. The two lakes were cut out from the felt with blue print underneath and silica gel on top of the acrylic for textural contrast. 

\subsection{Mini map of the Carlton Gardens}
\begin{figure}
\centering
  \includegraphics[width=0.9\columnwidth]{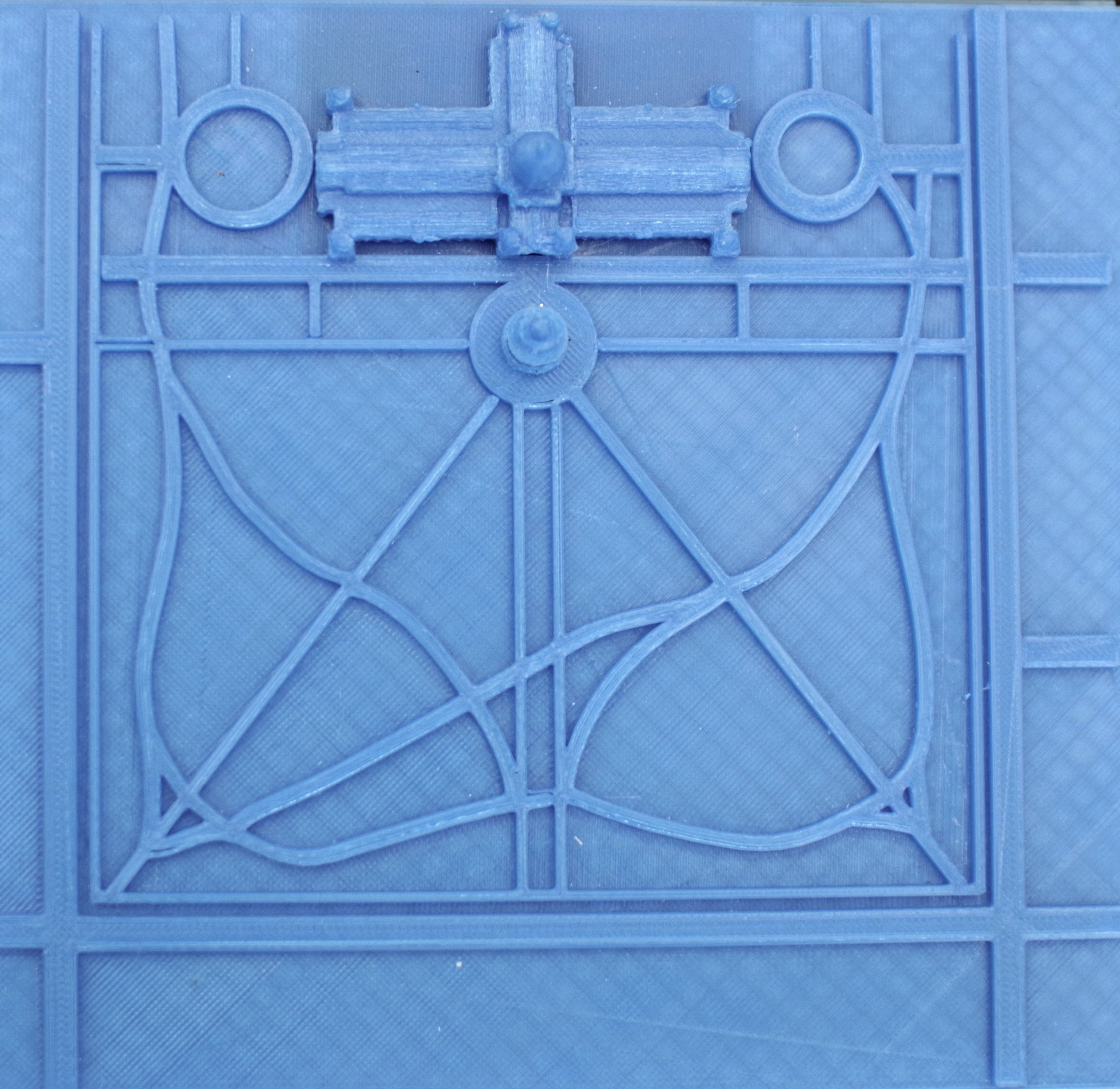}
  \caption{3D printed mini map with roads, paths, building and fountain.}~\label{fig:figure4}
\end{figure}

A handheld map of the gardens was also created at the much smaller size of 142mm $\times$ 128mm. 
This map was entirely 3D printed and included only a base (2mm thick), roads and pathways, the Exhibition Building and the fountain. The roads and pathways were raised (by 1.5mm) instead of lowered because they needed to be quite thin and would be difficult to perceive if indented.  

\subsection{Model of the Guide Dogs Victoria Sensory Garden}
\begin{figure}
\centering
  \includegraphics[width=0.9\columnwidth]{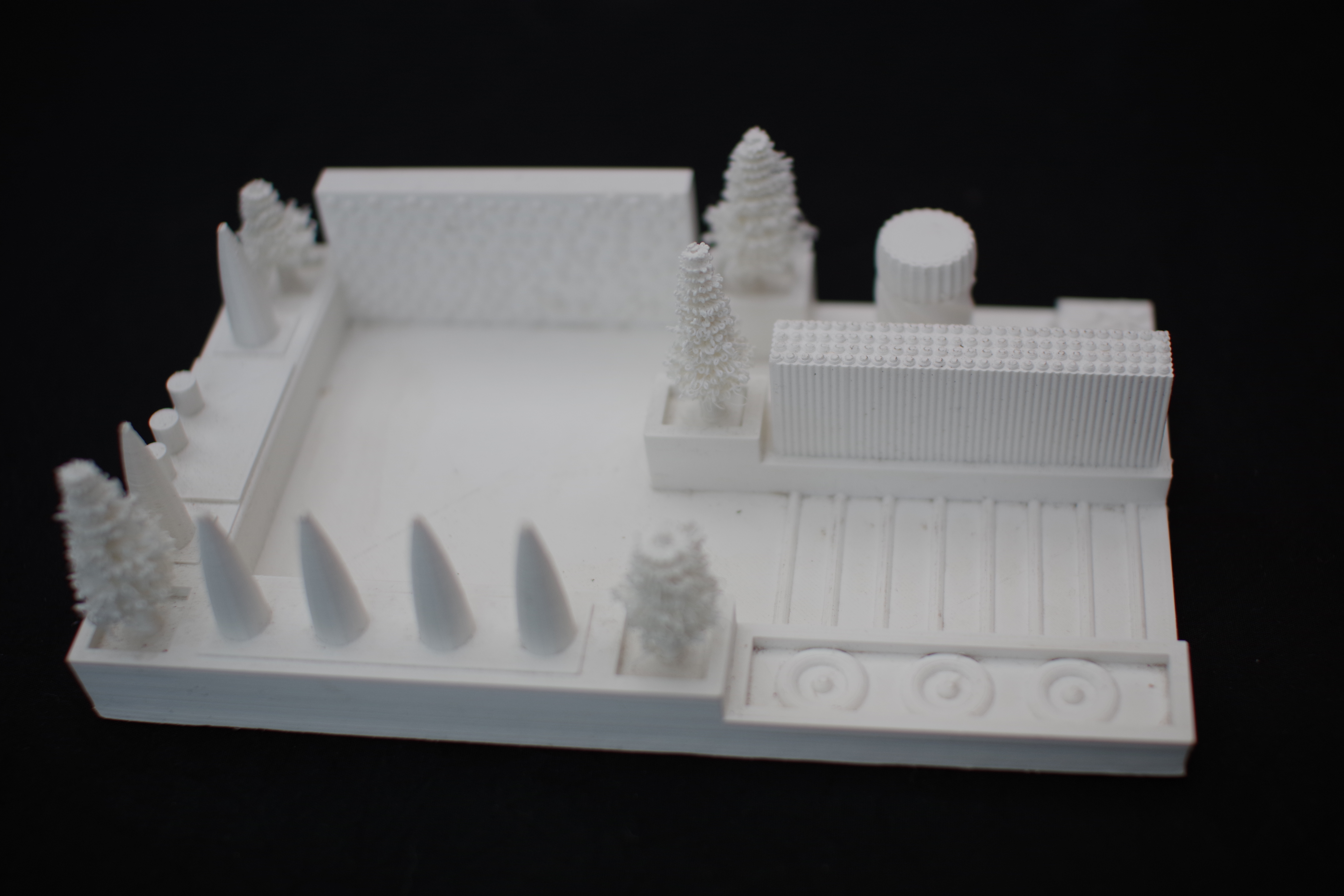}
  \caption{3D printed model of the Guide Dogs Victoria Sensory Garden.}~\label{fig:figure5}
\end{figure}

A 3D model for the GDV Sensory Garden was created in TinkerCAD based on a design sketch, photographs taken during installation and verbal advice from the project manager. As far as possible, measurements were followed to scale. The only allowance made for touch access was the omission of the marquee (roof) over a section of the garden. A variety of different textures were incorporated into the model to differentiate ground surfaces (wood versus plain), vegetation (herbs, reeds, flower beds, trees) and a water feature. 

\subsection{Model of the Rob Waddell Show Garden}
\begin{figure}
\centering
  \includegraphics[width=0.9\columnwidth]{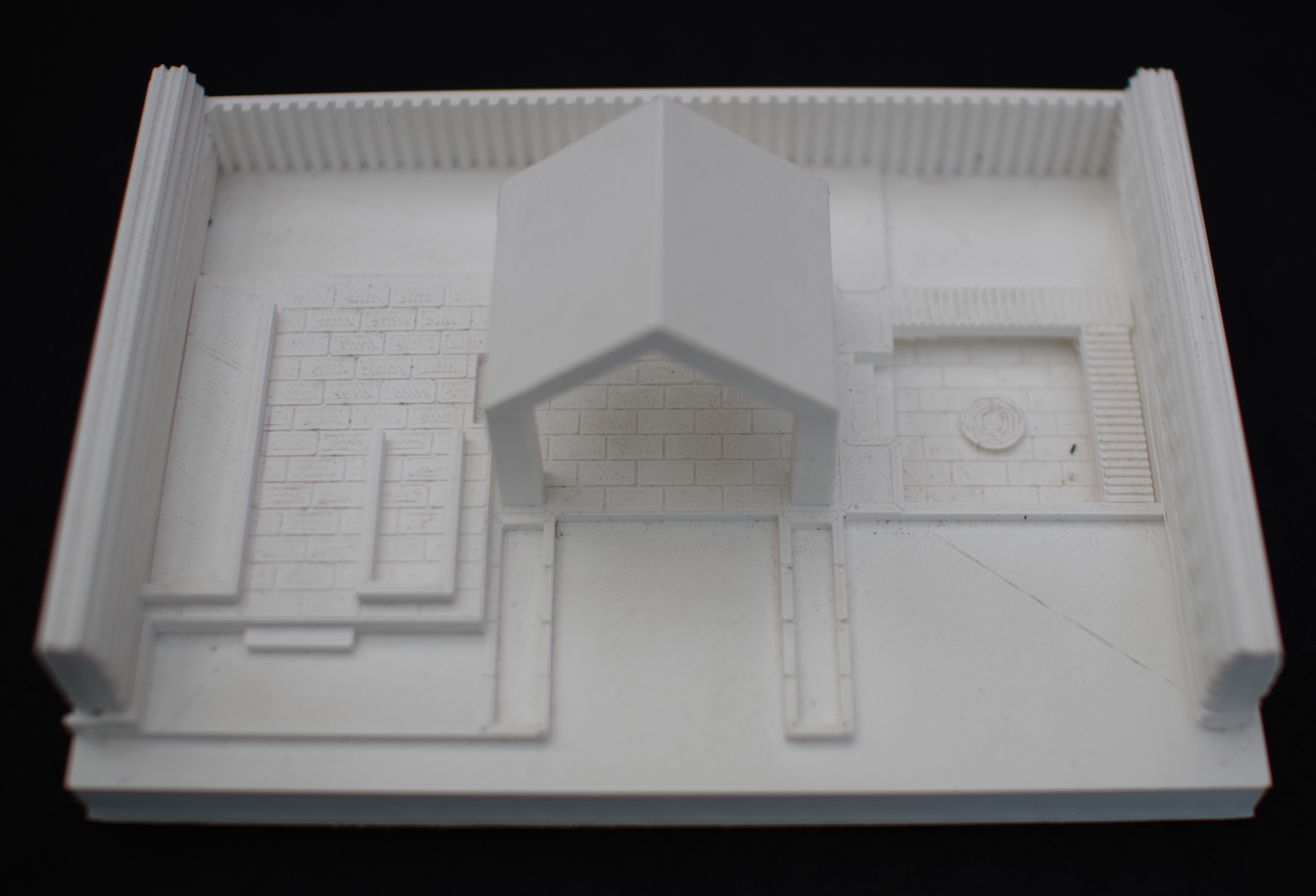}
  \caption{3D printed model of the Rob Waddell Show Garden. }~\label{fig:figure6}
\end{figure}

The final map depicted the Rob Waddell Show Garden. The 3D model was created by students at Kangan Institute with minimal guidance on design for tactile reading. Textures were included on the walls and paved area but the brickwork was difficult to perceive by touch as the pattern was very fine and indented. Due to practical consideration of timelines and the Ultimaker printing area, the dimensions were restricted to 180mm long  $\times$  122.5mm wide $\times$ 41mm high (plus roof). 

\section{Evaluation}

Our collaboration with Guide Dogs Victoria provided an opportunity to evaluate the usefulness of these maps for BLV visitors ``in the wild'' at the Melbourne International Flower and Garden Show.

\subsection{Procedure}
All maps were placed on display at the GDV Sensory Garden for the duration of the Garden Show. Unfortunately, practical constraints meant that the maps were placed on a low table in a crowded area near the sales and promotion tables at the rear of the garden. Training for GDV staff and volunteers working at the Garden Show included instruction to highlight the availability of the maps to all visitors. The research staff were on site and approached BLV visitors after they had looked at the maps to request their feedback. A series of set questions were asked over 10--30 minute semi-structured interviews. Responses were given as yes/no/maybe and additional comments were gathered via audio recording. The feedback was collected on site on the day (n=6), or shortly afterwards via phone (n=3) or at home (n=1). All participants looked at the large map of the Carlton Gardens but did not necessarily answer all questions or look at all three of the other maps, depending on their available time.

After the conclusion of the Garden Show, an email was sent to staff and volunteers who worked there asking for feedback via an online survey.

\subsection{Participants}
Guide Dogs Victoria had distributed 20 free tickets to the Garden Show to their clients and families. A total of ten BLV people provided feedback on the maps: Three were totally blind, six were legally blind with some vision, and one had low vision. Of the ten participants, seven were female. Ages ranged from 29 to 84 years, with an average age of 58.6 years.

Half of the BLV participants were braille readers, half could access enlarged print, and nine out of ten used audio to access print. Two were able to use standard print but these participants preferred other formats. Familiarity with tactile graphics ranged from none (n=2) and limited exposure (n=5) to substantial use (n=3). 

Most BLV participants used a range of strategies to assist with orientation and mobility. These included sighted guides (n=9), long cane (n=8), O\&M training to learn new areas (n=8), dog guides (n=7), mobile phone apps or other technology (n=5) and echolocation (n=1). One participant did not use any aids for orientation and mobility but they did report significant difficulties navigating hazards and crowds.

Six of the ten BLV participants had been to the Carlton Gardens before the Garden Show. On average, they had spent 1.5 hours at the Garden Show prior to using the maps.

Four people working at the Garden Show completed the online survey: Two staff who spent multiple days there and two volunteers who were each there for a single four-hour shift. None of the respondents worked directly with BLV clients in their usual daily roles.    
\subsection{Analysis}
A thematic analysis of all comments by participants was conducted independently by the four researchers. These themes were then compared and an agreed set was formed. The comments were then classified according to this agreed set of themes by two of the researchers independently, compared and any conflicts resolved. 

\section{Results and discussion}
Participants' responses and comments could be categorised into three main areas: Design guidelines, i.e. what features of the maps worked well and suggestions for improvements; application for orientation and mobility, i.e. how they could use the maps to learn about the depicted area and apply this knowledge for independent movement through the space; and  implementation, i.e. how such maps should be presented and some more general benefits that they can offer.

\subsection{Design guidelines (RQ1)}
\subsubsection{recognition of 3D printed icons}
Some icons were distinctive and could be recognised easily without reference to the key. Eight out of nine respondents said that all of the 3D icons felt different from one another. Every BLV participant in the study was able to instantly recognise at least one of the 3D printed icons, including those people who had never encountered a tactile map and one with age-related tactile desensitivity. Other icons were tactually distinct but required explanation or reference to the key. 

Participants were asked which icons were clearest and which were the most difficult to recognise. As seen in Table 1, the bowl with chopsticks representing food was easily recognised by eight of the nine people who mentioned it. The toilets, drinks, and fountain were also mentioned by a lot of people and found to be clear by the majority. By contrast, other icons were difficult to find (information), recognise by touch (water tap) or interpret (you are here). The water tap caused most confusion because it was felt from the top but there were important details lower down. 
\begin{adjustwidth}{1cm}{}
    ``You don't really feel down the side to the bottom. 
    You just feel the thing on the top.''
\end{adjustwidth}

\begin{table}
  \centering
  \begin{tabular}{l r r r r}
    icon & clear & difficult & \% favourable\\
    \midrule
    \underline{3D realistic}\\
    drinks & 4 & 3 & 57\%\\
    food & 8 & 1 & 88\%\\
    fountain & 3 & 2 & 60\%\\
    ticket booth & 1 & 1 & 50\%\\
    toilets & 5 & 3 & 62\%\\
    water & 0 & 6 & 0\%\\
    you are here & 3 & 4 & 43\%\\
    \underline{2.5D symbolic}\\
    first aid & 3 & 2 & 60\%\\
    information & 1 & 3 & 25\%\\
  \end{tabular}
  \caption{Number of BLV people who named each icon when asked which were clearest and which were most difficult to recognise.}~\label{tab:table1}
\end{table}

These results provide important evidence that 3D printed icons can be read and interpreted by people who are blind or low vision, even those who have little or no experience reading tactile graphics. However, not all of the icons tested were successful. Further work is required for the design of a set of easily recognised icons that can be shared among 3D map designers and learned by touch readers. 
\subsubsection{roads, pathways and textures}
Eight out of ten respondents said that the roads and pathways were clear and easy to follow. However, one BLV respondent reported that the pathways were confusing. This may be simply due to the complex nature of the pathways, however it was also noted that the 7mm wide indentation for pathways did not allow access with the full finger pad. Wider pathways may have been preferable. 
The recessed lakes were successful, with four people naming them as among the clearest things on the map. 

\subsubsection{realism versus symbolism}
When talking about the features on the maps and models, BLV participants expressed a desire for realism (14 comments by 5 people). The use of icons did cause some confusion regarding scale, however there was acknowledgement that such icons could be easier to understand than realistic features. 
\begin{adjustwidth}{1cm}{}
    ``This indicates where the public toilets are, is that right?  If you'd just done a little shed or something I wouldn't have known.''
\end{adjustwidth}

Likewise, it was important to people that the elements on the map be accurate in terms of position and scale (4 comments by 3 people).

\subsubsection{labelling}
Labelling or a key is clearly required because not all elements can be easily recognised by all people. The braille/print key was appreciated but could not be read by all people. It was suggested that the print labels be larger than N21 (2 comments by 2 people) and that audio labels could also be added. 

\subsubsection{standardisation}
BLV visitors suggested that the icons and symbols used on the maps should be standardised for easy recognition (4 comments by 3 people).
This same suggestion has been made in relation to tactile graphics but without sustained success due to the small number of symbols that can be recognised tactually compared with the large number of varied symbols used on print maps~\cite{Eriksson1998}. The use of more distinctive 3D icons may assist in addressing this problem but is unlikely to solve it altogether.  It is also noted that some of the successful 3D icons in this study were culturally specific and may not be suitable in other contexts. 

\subsubsection{scale}
With the addition of the vertical features, the mode in which 3D maps are explored tactually differs somewhat from 2.5D tactile graphics~\cite{Holloway2018printed}. Required sizes and gaps may also differ but have not been formally explored to date.  

The size of the maps and models was considered appropriate (3 comments by 3 people). None of the BLV participants complained about clutter, indicating that the height of icons (maximum 4cm) and spacing between them (minimum 1cm) was appropriate for effective tactual exploration. Successful recognition of the icons indicates that a minimum width of 1.5cm is sufficient for representational 3D icons. As already noted, indented pathways were considered clear at a width of 7mm but wider pathways may have been better.  

\subsubsection{complexity}
The smaller 3D models of the Sensory Garden and  Show Garden required less time to understand than the large map, which may have been too complex (5 comments by 3 people). 

\begin{adjustwidth}{1cm}{}
    ``[the Show Garden model is] very distinct and it's cut down to the bone. Not too much stuff. I think it's great.''
\end{adjustwidth}

All six people who looked at the small models  said that they were useful. Five out of six said that they helped them to build a mental model of the site and four out of four said that the textures were meaningful. 

As recommended for tactile graphics~\cite{Gardiner2002, Gardner1996}, 3D printed maps should be kept as simple as possible, with only the most important features and hazards included. 

\subsubsection{contrast}
High contrast was needed for people with low vision (3 comments by 3 people). The use of meaningful colours also assisted with decoding (3 comments by 2 people). 

\subsubsection{construction issues}
The use of felt caused some difficulties with secure placement of the 3D-printed icons. In addition, two of the participants suggested that the felt did not feel enough like grass and that a rougher material (sandpaper or fake grass) should be used instead. We also suggest designing icons with a wide base for more secure adhesion. 

Similarly, some of the 3D printed icons could have been better designed for durability. Some of the trees broke by the end of the five-day Garden Show and one BLV participant expressed a reluctance to touch the models in case they broke. 

\subsection{Supporting orientation and mobility (RQ2)}
When exploring and speaking about the maps, BLV visitors demonstrated an understanding that could be used to assist in wayfinding. Comments were classified into themes aligning with the four components of wayfinding described by Lynch~\cite{Lynch1960}: Orientation, route decisions, mental mapping and closure (detecting and arriving at the right place, i.e. successful wayfinding). 

\subsubsection{orientation}
The first step towards understanding the maps was to orient themselves (4 comments by 3 people). 
\begin{adjustwidth}{1cm}{}
``Is that north at the top?  If I get it oriented correctly, I find it easier''
\end{adjustwidth}
The maps were presented in the same direction as the environment they depicted, however the researchers observed that users often moved the map or themselves so their personal entry point faced towards them.

Landmarks could also easily be recognised using the maps (5 comments by 4 people). One blind visitor received a phone call from a friend elsewhere at the Garden Show while exploring the map and was able to assist them in identifying their location using the map. 

\subsubsection{route recognition and route finding}
While exploring the map, some of the BLV visitors spontaneously traced out the route that they took to enter the Garden Show or pointed in the direction of landmarks in the real world as they were talking about corresponding items on the map (4 comments by 3 people). 

\begin{adjustwidth}{1cm}{}
    ``We walked over grass then we walked over this way, that's the driveway there.''
\end{adjustwidth}

Further, six out of nine BLV respondents (67\%) said that they could use the map to help plan their visit, including the route. 

\subsubsection{mental model}
Seven out of ten BLV respondents (70\%) said the map helped them to imagine the gardens.   
\begin{adjustwidth}{1cm}{}
    ``I could have followed that in my head.  I had a picture of it from that map.''
\end{adjustwidth}

The BLV participants also spoke about the value of being able to form a mental model from the maps (6 comments by 6 people). 
\begin{adjustwidth}{1cm}{}
``If you get it mapped in your head, it's really easy.''
\end{adjustwidth}

\subsubsection{wayfinding}
Some participants demonstrated or believed that the map could assist with wayfinding (3 comments by 3 people). However, the map or mental model would need to be used in conjunction with other wayfinding strategies (2 comments by 2 people). 

\subsubsection{independence}
Most (80\%) of the BLV participants were at the Garden Show with family and friends,  
and spoke about relying on them for wayfinding (4 comments by 3 people). 
However, there was a desire for greater independence that could be achieved through use of accessible maps (1 comment by 1 person).
\begin{adjustwidth}{1cm}{}
    ``You go into [public] loos and you don't know whether to turn to the right, or the left, or where the cubicles are. It [a map] just would be so helpful. ... sometimes you go in on your own and oh god, it's a bit awful.'' 
\end{adjustwidth}

\subsection{Engagement and inclusion (RQ2)}
The 3D printed maps were certainly appealing to BLV and sighted visitors alike. 
Eight out of nine BLV respondents (89\%) said that the map was engaging (19 comments by 8 people).
\begin{adjustwidth}{1cm}{}
    ``Once you started feeling it, you wanted to feel more.''
\end{adjustwidth}

The researchers observed that the large map of the Garden Show was of great interest to sighted visitors. A number of sighted people used the 3D map to locate points of interest, even though the print map was on prominent display in the gardens and the event brochure. Children seemed particularly interested in the 3D map and parents used this interest as an opportunity to teach their children about map reading, accessibility and braille. 

These observations were supported by survey responses from the GDV representatives. All four respondents agreed that the 3D maps were of interest to sighted visitors, and observed that sighted visitors definitely (n=2) or maybe (n=2) used the accessible map to find points of interest in the garden. All four agreed that the maps conveyed a positive message about accessibility and inclusion. 

Application of the accessible maps ``in the wild'' demonstrated that, unlike tactile graphics, 3D printed maps are a format that can provide not just accessibility but also inclusion.

\subsection{Implementation of 3D printed maps (RQ2)}
\subsubsection{when and where}
The busy environment in which the maps were presented was far from ideal. The Garden Show event was too busy to allow for careful study of the large map and it needed to be placed in a more comfortable position -- preferably on an angle on a table with a chair available (5 comments by 3 people). Also, the large map required more time to study than most people had available. They estimated that around 30 minutes would be required (12 comments by 6 people). 

Particularly for the large complex map, participants expressed a strong desire to be able to study the map prior to visiting the location (5 comments by 4 people). 
\begin{adjustwidth}{1cm}{}
    ``I would have liked to have it before  so that I could really study it.''
\end{adjustwidth}
The maps could be provided at home, at a blindness agency or in a central location. In our example, the maps could have been on display at GDV headquarters prior to the event and situated at the two entry gates for the duration of the event, protected from the weather and on a table with chair. 

Others wanted a smaller map that they could carry with them (6 comments by 4 people). Only two of the participants took a copy of the mini-map with them while exploring the Garden Show, however many refused simply because they were not sure that they would have time to return it. Of those two people, only one used it. They reported that the minimap was useful, however further testing is required to evaluate the usefulness of handheld maps without the requirement that they be returned. 

\subsubsection{support required}
Two of the participants expressed a lack of confidence because they were not experienced in reading tactile maps (5 comments by 2 people). 

\begin{adjustwidth}{1cm}{}
    ``because I have such minimal experience,  it was a bit confronting''
\end{adjustwidth}

Nevertheless, these two participants were still able to  recognise the routes they had taken and felt that the maps had given them a good overview. These findings are particularly positive given that effective use of tactile graphics usually depends upon training and experience~\cite{Holloway2016}, 
but support and familiarisation with 3D maps will still be of assistance for some. 

It should be noted that one of the four GDV representatives indicated that they did not feel comfortable explaining the 3D printed maps to visitors. The researchers observed that few of the staff and volunteers were proactive in demonstrating the maps during the busy event. 

\subsubsection{suitable locations for mapping}
All four GDV representatives agreed that they would like to see 3D printed maps used by their organisation for accessibility in the future. Along with the BLV participants, they named numerous other sites where they would like to see similar 3D printed maps. Suggestions included shopping centres or shopping strips (6 comments by 5 people), public transport (3 comments by 3 people), the city precinct (3 comments by 2 people), internal maps of public places (2 comments by 2 people), and other public spaces (9 comments by 5 people). They suggested that maps of this sort would be most useful for permanent public spaces that will be used by a lot of people over a long time period (2 comments by 2 people). It may not be worth studying a map to form a mental model of a temporary event space such as the  Garden Show. 
\begin{adjustwidth}{1cm}{}
    ``You don't put too much time and effort into studying something that is of a temporary nature. But if it is a more or less permanent fixture, then you do sit down and study it very carefully.''
\end{adjustwidth}

\section{Conclusions}

In collaboration with Guide Dogs Victoria, we evaluated the use of 3D printed maps by BLV and sighted visitors at a large public event. The study provided valuable input from a natural sample of accessible map users ``in the wild''. While clearly relevant to RQ2, testing ``in the wild'' also gives greater ecological validity to the findings for RQ1. Interaction with the maps was affected by the busy noisy environment, planned visit duration, weather and so on. For example, most participants tried to interpret the large map as quickly as possible, only referring to the key when needed. Insights were also gained because participants were in the mapped environment and motivated to explore the physical area. They spontaneously looked for their current location, found their entry point, and traced the routes they had walked or planned to travel. Testing ``in the wild'' also provided access to a representative population sample, including people who may not volunteer for more formal research because they are not connected with blindness agencies, have recent vision loss, or are not confident reading tactually. 

Based on this stakeholder feedback, we were able to develop the following recommendations for the design and use of 3D printed maps for accessibility.
\subsection{Design Guidelines (RQ1)}
\begin{itemize}
  \setlength\itemsep{0em}
\item Representational 3D icons can be used in many cases and hold the following advantages over abstract symbols:
\begin{itemize}
  \setlength\itemsep{0em}
  \item They are more inclusive, allowing sighted and vision impaired people to use the same map, and enabling access by BLV people who cannot read braille or large print. 
  \item They are more enjoyable and engaging than abstract symbols.
  \item Cognitive demand is reduced, as many icons can be understood without reference to a legend. 
\end{itemize}
\item Representational 3D icons can be recognised tactually at a minimum size of approximately 1.5cm$^{3}$, with at least 1cm gap between adjacent icons.
\item 3D icons placed on a base map should have the most important features at the top for easy access by touch.
\item Indented pathways work well for large maps, however they should be more than 7mm wide. 
\item For maximum accessibility and inclusion, maps should consider a range of accessibility preferences, making use of: 
\begin{itemize}
  \setlength\itemsep{0em}
    \item large print and high contrast visuals, including meaningful use of colour
    \item a range of differing textures
    \item a key with print, braille and audio
\end{itemize}
\end{itemize}

\subsection{Implementation Guidelines (RQ2)}
\begin{itemize}
  \setlength\itemsep{0em}
\item Create 3D accessible maps of high-use public spaces.
\item Training and experience is not necessary to read and understand accessible 3D maps, however familiarisation with commonly used icons would be helpful. 
\item Maps should ideally be made available for study before going to the mapped location. They should also be provided at entrances.
\item Large, complex maps should be tilted for use while seated as they can require considerable time to explore and understand. The map and reader should be protected from weather.
\item Accessible 3D maps provide a positive message about inclusion and  accessibility, as they are of interest and use to the sighted community as well as people who are blind or have low vision.
\end{itemize}

\subsection{Future Research}
The usefulness of portable maps for route finding could not be properly evaluated due to the requirement that the maps be returned and time constraints of participants. This is an area for future investigation. Further, interesting questions were raised about mixing symbolic and representational icons, which could benefit from further exploration.

\section{Acknowledgments}

Our sincere thanks are extended to the staff, volunteers and visitors to the Garden Show who assisted in the creation and use of the 3D maps. 

We gratefully acknowledge the Australian Research Council and our project partners Guide Dogs Victoria, Round Table on Information Access for People with Print Disabilities Inc., the Department of Education and Training Victoria, the Royal Institute for Deaf and Blind Children, and the Royal Society for the Blind. 

Thanks also to the paper reviewers for their helpful questions and suggestions. 

\balance{}

\bibliographystyle{SIGCHI-Reference-Format}
\bibliography{main}

\end{document}